\documentclass[twocolumn,showpacs,preprintnumbers,amsmath,amssymb]{revtex4}

\usepackage{graphicx}
\usepackage{dcolumn}
\usepackage{bm}

\begin{document}

\preprint{APS/123-QED}

\title{ Homogeneous coexistence of SDW and SC states in CaFe$_{1-x}$Co$_{x}$AsF studied by nuclear magnetic resonance }

 \author{ T. Nakano$^1$, S. Tsutsumi$^1$, N. Fujiwara $^1$ \footnote { Corresponding author: naoki@fujiwara.h.kyoto-u.ac.jp},
S. Matsuishi$^{2, 3}$, and H. Hosono$^{2, 3}$
 }

\affiliation{$^1$ Graduate School of Human and Environmental
Studies, Kyoto University, Yoshida-Nihonmatsu-cyo, Sakyo-ku, Kyoto
606-8501, Japan}


\affiliation {$^2$ Materials and Structures Laboratory (MSL), Tokyo
Institute of Technology, 4259 Nagatsuda,  Midori-ku, Yokohama
226-8503, Japan \ \\ $^3$ Frontier Research Center (FRC), Tokyo
Institute of Technology, 4259 Nagatsuda, Midori-ku, Yokohama
226-8503, Japan}




\date{November 30 2010}


\begin{abstract}

 We investigated the homogeneous coexistence of spin-density-wave
 (SDW) and superconducting (SC) states via $^{75}$As-nuclear magnetic resonance (NMR) in
 CaFe$_{1-x}$Co$_{x}$AsF and found that the electronic
 and magnetic properties of this compound are intermediate between those of LaFeAsO$_{1-x}$F$_x$ and
 Ba(Fe$_{1-x}$Co$_x$)$_2$As$_2$. For
 6\% Co-doped samples, the paramagnetic spectral weight completely disappears in the
crossover regime between the SDW and SC phases followed by anomalous behavior of relaxation rate ($1/T_1$), implying that the two
 phases are not segregated. $^{59}$Co-NMR spectra
show that spin moments are not commensurate but spatially modulated.
These experimental results suggest that
 incommensurate SDW (IC-SDW) and SC states are compatible in this
 compound.

\end{abstract}

\pacs{74.70. Xa, 74.25. Dw, 74.25. nj, 76.60. -k}
\maketitle

To date, a number of iron-based high-critical temperature ($T_c$)
superconductors have been discovered since the first discovery of
superconductivity in LaFeAsO$_{1-x}$F$_x$ [1]. Among these, 1111
series (RFeAsO$_{1-x}$F$_x$ [R=Nd, Sm, Ce, and La, $etc.$) [1-6] and
122 series such as Ba(Fe$_{1-x}$Co$_x$)$_2$As$_2$ [7-9] have been
studied intensively because the former has a relatively high $T_c$
compared with the others, and the latter is available in a
single-crystal form. A clear difference between the two series is
apparent from the electronic phase diagrams shown in Fig. 1. The
structural and magnetic phase transitions occur at almost the same
temperature ($T$) for the 122 series but not for the 1111 series.
The optimal $T_c$ for the La1111 series is realized away from the
antiferromagnetic (AF) or spin-density-wave (SDW) phase, whereas the
optimal $T_c$ for the Ba122 series is realized near the AF phase,
although the optimal $T_c$ values are almost the same at ambient
pressure. In other words, the overlap between the AF and SC phases
is large for the Ba122 series, whereas the two phases are almost
separated in the La1111 series. This necessarily leads to the
problem of coexistence in the crossover regime. Several theoretical
investigations have focused on charge and spin states in the
vicinity of the crossover regime [10-12]. The coexistence of the
order parameters of the incommensurate spin-density-wave and
superconducting (IC-SDW + SC) states has been investigated in the
crossover regime for s$^{+-}$- or s$^{++}$-wave symmetry [10-12].
The homogeneous coexistence of the AF and SC states has been
suggested for Ba(Fe$_{1-x}$Co$_x$)$_2$As$_2$ on a microscopic level
[13-15], whereas microscopic or mesoscopic segregation of the AF and
SC phases has been suggested for K-doped
Ba(Fe$_{1-x}$Co$_x$)$_2$As$_2$ [16-18]. The topics have been
investigated mainly for the Ba 122 series [19] and it is presently
unclear whether the observed phenomena are inherent to all
iron-based superconductors with a crossover regime.

\begin{figure}
\includegraphics{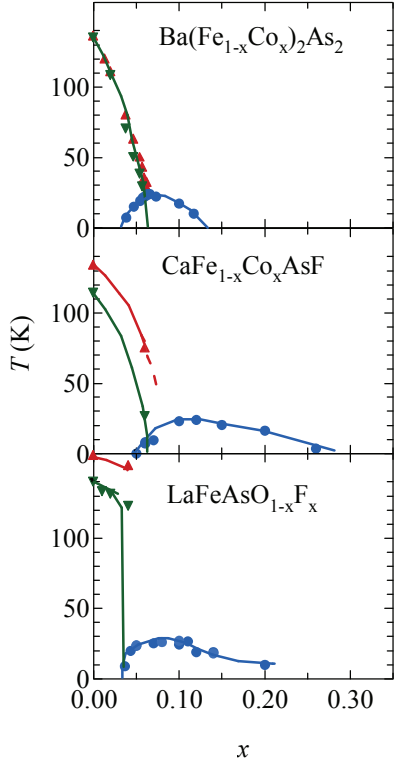}
\caption{\label{fig:epsart} (Color online) Phase diagrams of
Ba(Fe$_{1-x}$Co$_x$)$_2$As$_2$ [7-9], CaFe$_{1-x}$Co$_{x}$AsF
[20-23] and LaFeAsO$_{1-x}$F$_x$ [1, 2]. Red and green closed
triangles and blue circles represent the structural ($T_S$), AF
($T_N$) and SC ($T_c$) transition temperatures, respectively. }
\end{figure}

In this study, we focus on CaFe$_{1-x}$Co$_{x}$AsF [20] to
investigate whether the AF and SC phases coexist homogeneously or
are segregated.
 The compound is a member of the 1111 series, but exhibits an intermediate phase diagram, as shown in Fig.
1 [20-23]. The degree of overlap of the SC and AF phases is
intermediate between the La1111 and Ba122 series. The doping
dependence of the structural ($T_S$ ) and magnetic ($T_N$)
transition temperatures is also intermediate between the two series.

We performed nuclear magnetic resonance(NMR) measurements using 6\% and 12\% Co-doped powder
samples. Figures 2 (a) and (b) show the $T$ dependence of the
resistivity. The SC phase was confirmed for both samples; the
$T_c$ onset is 7 and 20 K for the 6\% and 12\% Co-doped samples,
respectively. The detuning of an NMR tank circuit is shown in Fig. 2
(c). The bending point at 7 K is closer to the zero-resistivity
temperature than the $T_c$ onset.

\begin{figure}
\includegraphics{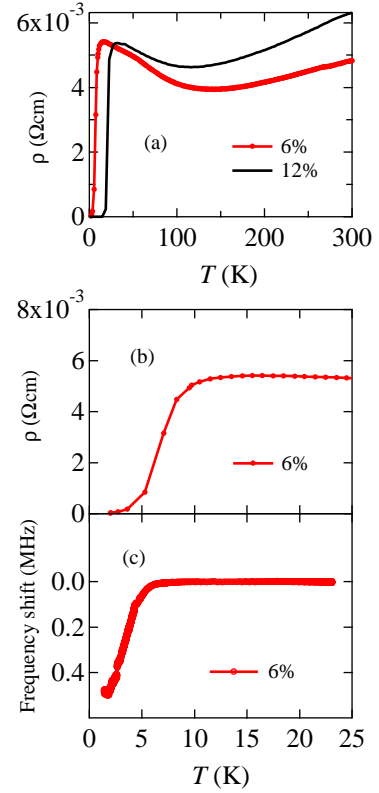}
\caption{\label{fig:epsart} (Color online) (a), (b) Superconducting
transition temperature determined by resistivity and (c) detuning of
an NMR tank circuit. Figure. 2 (b) is an expanded view of Fig. 2
(a). These measurements were operated at zero field.}
\end{figure}

For the 6\% Co-doped samples, the field-swept $^{75}$As-NMR spectra
were measured at a frequency $\nu_0$ = 45.1 MHz, as shown in Fig. 3.
The low-field peak around 50 kOe corresponds to the transition I =
-3/2 $\Leftrightarrow$ -1/2, and the central peak with two bumps
corresponds to the transition I = -1/2 $\Leftrightarrow$ 1/2. The
two bumps of the central peak originate from the second-order
quadrupole effect. The frequency shift ($\nu$) arising from this
effect is expressed using two polar coordinates, $\theta$ and
$\phi$, of the applied field ($\emph{\textbf{H }}$ ) in the
coordinate system of the electric field gradient (EFG). The
coordinate $\theta$ represents the angle between $\emph{\textbf{H
}}$ and the maximal principal axis (Z axis) of EFG. In this case,
the Z axis is equivalent to the crystal c axis. The coordinate
$\phi$ represents the angle between the second principal axis (X
axis) and the $\emph{\textbf{H }}$component projected onto the XY
plane. The XY plane coincides with the crystal ab plane. The
$\theta$ and $\phi$ dependence of $\nu$ is given as
\begin {equation}
\nu(\theta, \phi) = - \frac{3 \nu_Q^2}{16 \nu_0}{sin^2
\theta}[(9cos^2 \theta-1) + \eta f(\theta, \phi)+ \eta^2 g(\theta,
\phi)],
\end {equation} where $\nu_Q$ and $\eta$ ($0 < \eta < 1$) are the pure quadrupole frequency and
 the EFG anisotropy, respectively. From the
analysis of the NMR spectra, $\nu_Q$ and $\eta$ are estimated to be
22.6 MHz and 0.3, respectively.  The value of $\nu_Q$ is almost
twice as large as that for the La1111 series [24]. The terms $\eta
f(\theta, \phi)$ and $\eta^2 g(\theta, \phi)$ in Eq. (1) are
correction terms arising from axial asymmetry.
 The $\theta$ and $\phi$ dependence of $f$ and $g$
is explained in detail [25]. At the field $h=H-H_0$, where $H_0$
represents the resonance field expected for $\nu_Q = 0$, the
spectral intensity $I^{As} (h)$ is
\begin {equation}
   I^{As}(h) \propto \int \int \delta ( h - (2\pi / \gamma_N)\nu(\theta, \phi)) d(cos \theta
  )d\phi,
\end {equation} where the gyromagnetic ratio ($\gamma_N$) of $^{75}$As is 7.292 MHz/10kOe.
   For $\eta = 0$, \begin {equation}
   I^{As}(h) \propto \int  \delta (h - \frac{2\pi}{\gamma_N} \nu) |\frac{sin
\theta}{\frac{\partial \nu}{\partial\theta}}|d\nu.
\end {equation} The bumps originate from the powders with $\theta$ = 90$^{\circ}$ or
42$^{\circ}$ that satisfy $ \frac{\partial \nu}{\partial\theta} /
sin \theta = 0$. Even for $\eta \neq  0$, the bumps appear at the
same angle ($\theta$ = 90$^{\circ}$ or 42$^{\circ}$ ) [25].
 The low-field bump originates from the powders with $\theta$ =
90$^{\circ}$ in which the Fe basal planes are aligned with
$\emph{\textbf{H}}$, whereas the high-field bump originates from the
powders with $\theta$ = 42$^{\circ}$. The spectral weight weakens
below $T_N$ and completely disappears in the SC state. An analogous
disappearance is not observed for the 12\% Co-doped samples in which
the AF phase is absent. The disappearance of the bump is
characteristic of the AF phase.

\begin{figure}
\includegraphics{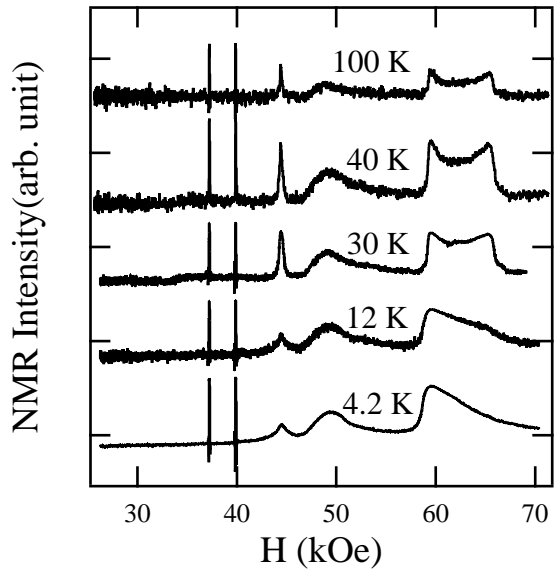}
\caption{\label{fig:epsart} Field-swept $^{75}As$-NMR spectra at
several temperatures for the 6\%-Co doped samples. The sharp signals
at 37.3 and 40.0 kOe originate from $^{65}$Cu and $^{63}$Cu,
respectively, which are present in the copper NMR coil. The
cusp-shaped signal observed at 45 kOe originates from $^{59}$Co in
the Fe basal planes. The other signals originate from $^{75}$As. The
peak around 50 kOe corresponds to the transition I = -3/2
$\Leftrightarrow$ -1/2, and the double-bump peak corresponds to I =
-1/2 $\Leftrightarrow$ 1/2. }
\end{figure}

  Unlike the
high-field bump, the low-field bump is maintained even at low
temperatures. The resonance position of the low-field bump is almost
unchanged even when the Co nuclei experience the internal field that
appears in the AF phase. For the commensurate SDW state, stripe-type
antiferromagnetic spin moments aligned in an Fe plane (the ab plane)
induce the internal field normal to the Fe plane on $^{75}$As sites,
therefore, As nuclei experience no internal field when
 the Fe plane is parallel to $\emph{\textbf{H }}$($\theta$ = 90$^{\circ}$).
Consequently, the low-field bump is maintained even in the
spin-ordered state. For crystals with other $\theta$ angles, As
nuclei experience nonzero internal fields, and the corresponding
signals are swept over a wider field range. This explanation is
possible for the IC-SDW state accompanied by
 long-wave spin modulation.

\begin{figure*}
\includegraphics{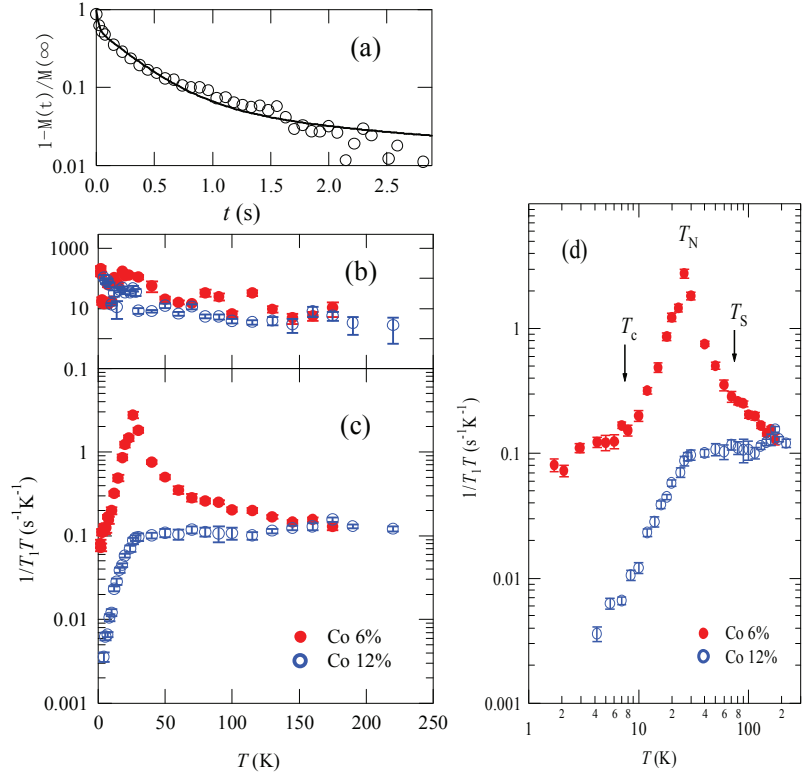}
\caption{\label{fig:wide} (Color online)(a) Recovery curve of the
central transition (1/2 $\Leftrightarrow$ -1/2) measured at 4.2 K
for the 6\% Co-doped samples. The data are fitted using Eq. (4).
(b), (c) Temperature dependence of $1/T_{1s}T$ and $1/T_{1}T$ for
the 6\% and 12\% Co-doped samples. (d) Logarithmic plot of Fig. 4
(c). The bending point at 7 K coincides with the transition
temperature ($T_c$). }
\end{figure*}

The relaxation time $T_1$ was measured using the conventional
saturation-recovery method.
 The $T$ dependence of  $1/T_1T$ was measured at the low-field bump. The bump position is almost unchanged for the observed temperatures.
 Therefore, $1/T_1T$ is successively measured even in the AF phase with the same conditions as the paramagnetic phase. The recovery curves $1-M(t)/M(\infty)$, where $M(t)$ represents
 the nuclear magnetization after the saturation pulse, exhibit a fast-recovery component throughout
 the entire
 temperature range. The data were analyzed using three-component
 fitting; that is, fast-recovery component $1/T_{1s}$ is introduced tentatively into the recovery curve as
 follows:
\begin{equation}
1-\frac{M(t)}{M(\infty)}=c(0.9e^{-6t/T_1}+0.1e^{-t/T_1})+(1-c)e^{-t/T_{1s}}.
\end{equation}

A recovery curve measured at 4.2 K for the 6\% Co-doped samples is
shown in Fig. 4 (a). The magnitude of $1/T_{1s}$ is two orders
larger than that of $1/T_1$, which allows us to separate the two
terms without ambiguity. The $T$ dependence of $1/T_{1s}T$ and
$1/T_{1}T$ is shown in Figs. 4 (b) and 4 (c), respectively. Below 50
K, the value of $c$ is approximately 0.4 to 0.5. The results of
$1/T_1T$ are almost unchanged even if the experimental data are
analyzed using $0.9e^{-t/6T_{1s} } + 0.1e^{-t/T_{1s}} $ instead of
$e^{-t/T_{1s}}$ in Eq. (4). The $T$ dependence of $1/T_{1}T$
reflects the AF and SC transitions both, whereas that of $1/T_{1s}T$
reflects neither of them. In Fig. 4 (d), $T_S$ =75 $\pm$ 5 K and
$T_N$ =26 K are derived from the bending point and peak,
respectively. From neutron scattering measurements, the magnitude of the
ordered moments is estimated to be 0.15
 $\mu_B$, which is considerably small compared with that in the undoped
 case (0.49
 $\mu_B$) [22].
   The bending point at 7 K observed in Fig. 4 (d) corresponds
  to $T_c$, which is consistent with the temperatures observed from the detuning of the tank circuit and the resistivity measurements.
   Although $1/T_1$ was measured under an applied magnetic field, $T_c$ is almost the same, implying that the decrease in $T_c$ due to the applied field is
negligible. As a result, the effect of the vortex is neglected when
the Fe basal planes are parallel to $\emph{\textbf{H}}$. Actually,
we could not observe the field dependence of $1/T_{1}$ which is
expected to appear for the vortex motion.
  The temperature dependence at low temperature follows the power law, $1/T_1 \sim T^{1.5}$. We tentatively attribute this dependence to an
intrinsic electronic state in the Fe basal planes. The fast-recovery
component  $1/T_{1s}T$ is also observed for the 12\% Co-doped samples.
   Therefore, this component may be extrinsic to the low-frequency fluctuation arising from the Fe basal planes.
   A possible origin of the fast-recovery component is nuclear
   spin diffusion.

 For the 12\% Co-doped samples,  $1/T_1T$ is almost independent of the temperature above $T_c$ and is proportional to $T^2$ below $T_c$. There are two possible origins for this temperature-independent
 behavior. The first possibility is a Korringa relation $1/T_1T\propto D(\varepsilon_F)^2$, where $D(\varepsilon_F)$ is the density of the states
 at the Fermi energy, and the second possibility is very weak Curie-Weiss behavior. Both possibilities have been observed in LaFeAsO$_{1-x}$F$_x$. The former
has been observed for the overdoped regime ($x$ = 0.14-0.15) just
above $T_c$ [24, 28], whereas the latter has been observed for the
underdoped regime in a wide temperature range above $T_c$ [24, 29].
In the present case, the value of $1/T_1T$ is rather close to that
observed for the underdoped regime. The $T$ dependence is
attributable to spin fluctuations.

\begin{figure*}
\includegraphics{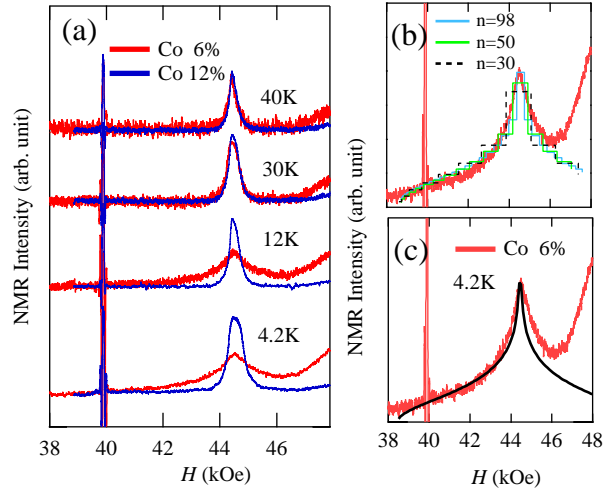}
\caption{\label{fig:epsart} (Color online) (a)$^{59}$Co-NMR spectra
at 45.1 MHz for the 6\% and 12\% Co-doped samples. The cusp-type
symmetric pattern for the 6\% Co-doped samples was unchanged even at
low NMR frequencies. (b) Simulation of the spectral intensity for
the internal field $\pm h_{AF}sin(Qr)$, where the wavelength is 2$n$
lattice units, $Q = \pi/n$ and $h_{AF}$ = 6 kOe. (c) Spectral
intensity calculated with $h_{AF} = 6$ kOe in Eq. (8).}
\end{figure*}

The present NMR measurements confirm successive phase transitions.
The complete disappearance of the high-field bump in the crossover
regime indicates that no paramagnetic domains exist, which implies
that no phase segregation occurs in this doping regime. The fact
that the AF phase is not commensurate can be demonstrated by the
$^{59}$Co($I=7/2$)-NMR spectra shown in Figs. 5(a) - (c). The
satellite signals expected owning to the quadrupole effect are not
clearly observed in the paramagnetic state as seen in Fig. 5(a),
implying that the EFG at Co nuclei is considerably small and is
somewhat distributed. The broadening due to an inhomogeneous EFG or
hyperfine coupling may be seen at low temperatures for the 12\%
Co-doped samples. For the 6\% Co-doped samples, however, the
inhomogeneity hardly explains the broadening unless some spin
ordering is considered. If a commensurate SDW state with an internal
field $\pm h_{AF}$ was realized in the crossover regime, the
spectral intensity would be
\begin{eqnarray}
 I^{Co}(h) &\propto& \int  \delta (h - h_{AF}cos \psi) d(cos
\psi) \\
&\propto& \frac{1}{h_{AF}} \   \ \   \ \   \ ( |h/h_{AF}| \leq 1),
\end{eqnarray} where $h = H-H_0$ and $\psi$ represents the
angle between $\emph{\textbf{H }}$ and the Fe plane. The field
dependence represents a well-known rectangular powder pattern,
implying that the commensurate SDW state is not realized. Figure
5(b) shows simulations for the internal field modulated to the AF
direction $\pm h_{AF}sin(Qr)$, where the wavelength is 2$n$ lattice
units, $Q = \pi/n$ and $h_{AF}$ = 6 kOe. The width of the top
rectangular pattern is proportional to $1/n$ for a large $n$. A
pattern with a large $n$ approaches to the curve expressed by the
following integral:
  \begin{eqnarray}
 I^{Co}(h) &\propto& \int_{sin^{-1}(h/h_{AF})}^{\pi/2}\frac{1}{h_{AF}sin\varphi}
 d\varphi \\
 &\propto& ln |\frac{1+\sqrt{1-(h/h_{AF})^2}}{h/h_{AF}}|. \   \ ( |h/h_{AF}| \leq 1)
\end{eqnarray} The curve reproduces the experimental results well as shown in Fig. 5(c). In fact, an extremely large $n$ is
not always required owning to the small quadrupole effect: the
rectangular pattern would become smooth when this effect is
considered.The simulations for $n$ = 50 - 100 could reproduce the
experimental results. The cusp-type pattern at 4.2 K in Fig. 5(a) is
almost the same as that at 12 K, suggesting that the distribution of
the internal field below $T_c$ is identical to that above $T_c$.
This result implies that the SC transition proceeds on the
background of the IC-SDW state which is well developed above $T_c$.
A uniform electronic state, the IC-SDW + SC state, is a promising
candidate to explain these experimental results.

 In summary, we measured NMR spectra and
$1/T_1T$ in CaFe$_{1-x}$Co$_{x}$AsF and found that the electronic
 and magnetic properties are intermediate between those of LaFeAsO$_{1-x}$F$_x$ and
 Ba(Fe$_{1-x}$Co$_x$)$_2$As$_2$. For the 6\% Co doped samples, we confirm the coexistence of the SC and spin-ordered states from $1/T_1T$
 at low temperatures.
 At $\theta = 42 ^{\circ}$, the  $^{75}$As spectral weight  completely
disappears in the crossover regime, which indicates that no phase
segregation is realized in this regime. The $^{59}$Co-NMR spectra
demonstrate that spin moments are not commensurate but spatially modulated.
In addition, the SC state develops on the background of the IC-SDW state,
which suggests the existence of a uniform electronic IC-SDW + SC
state.







\end{document}